\documentclass[11pt]{article}

\usepackage{url}
\usepackage{graphicx}
\usepackage{amsmath}
\usepackage{amssymb}

\usepackage{makeidx}

\makeindex

\allowdisplaybreaks[2]

\begin{document}

% rename the file example.tex as appropriate
\section{Investigating Partons' Transverse Spin with Deep Inelastic Exclusive Experiments}
\label{sec:example}

%%%%%%%%% insert author names addresses here, please do not change format

\hspace{\parindent}\parbox{\textwidth}{\slshape 
  Gary R. Goldstein$\,^1$, Simonetta Liuti$\,^2$ \\[1ex]
$^1$\ Department of Physics and Astronomy, Tufts University, Medford, MA, USA \\
$^2$\ Department of Physics, University of Virginia, Charlottesville, VA,
  USA }

% put the author names here again, for the list of authors at the end of
% the volume
\index{Goldstein, Gary}
\index{Liuti, Simonetta}

\vspace{2\baselineskip}

%%%%%%%%%%%%%%%%%%%%%%%%%%%%%%%%%%%%%%%%%%

\centerline{\bf Abstract}
\vspace{0.7\baselineskip}
\parbox{0.9\textwidth}{We show how deeply virtual pseudoscalar meson production experiments single out the contribution of the four chiral 
odd quark-proton GPDs which are related to the so far elusive transversity distribution function, $h_1$. 
Furthermore, in the kinematical ranges of the proposed EIC, electroproduction of strange and charmed mesons will allow one to uniquely pin down the 
non-perturbative charmed component in the nucleon structure function, and 
at the same time provide new insights in the 
connection of the quark/gluon 
degrees of freedom with the meson-baryon description. 
The use of dispersion relations as well as the naive extension of the parton model to the ERBL region are critically analyzed.
}

%%%%%%%%%%%%%%%%%%%%%%%%%%%%%%%%%%%%%%%%%%

\subsection{Introduction}
%We report here on recently completed work~\cite{Ahmad:2008hp,Goldstein:2010gu} and work in progress. 
In this contribution we suggest a class of deeply virtual exclusive reactions, namely pseudoscalar meson electroproduction, as a 
means to access chiral odd distributions.
These are described by a set of four chiral odd GPDs which enter the matrix elements for the various terms of the cross section.  
We conducted an analysis using a parametrization of the GPDs that is inspired by a physically motivated picture of the nucleon as a quark-diquark system with Regge behavior. In the chiral even sector a quantitative parameterization can be obtained from a global fit to PDFs, nucleon form factors, and DVCS data where the masses, couplings and Regge power behavior that set the scale for the dependence on the kinematic variables, $X, \zeta, t, Q^2$, are determined via a {\em recursive} procedure~\cite{Goldstein:2010gu}. 
%In Ref. \cite{Goldstein:2010gu} we provide a parameterization by first fitting the parton distribution functions $f_1$ and $g_1$ for the u and d quarks with $H(X,0,0)$ and $\widetilde{H}(X,0,0)$ at a low scale. The electromagnetic form factors, $F_1(t)$ and $F_2(t)$ or the first $X$ moments of $H(X,\zeta, t)$ and $E(X,\zeta,t)$ are constrained to satisfy polynomiality, thereby removing the $\zeta$ dependence and leaving only the $t$ dependence. This requires fixing the parameterization of the ERBL region, $X<\zeta$, for all $\zeta$ so as to satisfy a sum rule for the form factor. The same approach is used for the axial vector and pseudoscalar  form factors constraining  
%$\widetilde{H}$  and $\widetilde{E}$, respectively. 

%{\bf write connection to chiral odd, Gary}
The extension of this parameterization scheme to the chiral odd GPDs is critical for the phenomenology of deeply virtual meson electroproduction, which was begun particularly for the $\pi^0$ in Ref.~\cite{Ahmad:2008hp}. In the diquark spectator model the incoming and outgoing quark-nucleon vertices communicate through the exchanged diquark. For the helicity amplitudes, then, these vertices factorize (for each helicity of the diquark, scalar or axial vector), so that chiral even helicity amplitudes are simply related to their chiral odd counterparts via parity transformations of one of the vertices. For the d-quark case it is only the axial diquark relations that are involved, while the u-quark involves the scalar  contribution, as well. We thereby obtain the full set of four  chiral odd GPDs, each being linearly related to helicity amplitudes. This allows us to predict the behavior of pseudoscalar electroproduction \cite{GGL_progress}. 

It has now become particularly pressing to study the heavy quark components of the nucleon because of the advent of the LHC.  
%The heavy quark components will be key in the study
%of QCD matrix elements in the unprecedented multi-TeV CM energy regimes accessible at the LHC.
%At the same time, as the LHC opens new horizons for studies of physics beyond the Standard Model,  many ``candidate theories'' will provide similar signatures of a departure from SM predictions.
For the types of precision measurements in the unprecedented multi-TeV CM energy regimes envisaged at the LHC it will be necessary to provide accurately determined QCD inputs.
The analyses in Ref.\cite{Pum} have shown how the inclusion of non perturbative charm quarks could modify the outcome of  global PDF analyses. However, the situation is not clear-cut. 
We therefore extended our analysis to strange and charm psuedoscalar meson production \cite{Liuti:2010xy}. 
We proposed that in order to refine analyses such as the one Ref. \cite{Pum} new observables are identified from deeply virtual meson production, and spin correlations measurements. 
We presented preliminary results involving the following electroproduction exclusive processes: 
(1) $\gamma^* p \rightarrow J/\psi \, p^\prime$; (2) $\gamma^* p \rightarrow D \, \overline{D} \, p^\prime$; (3) $\gamma^* p \rightarrow \overline{D} \, \Lambda_c$; 
(4) $\gamma^* p \rightarrow \eta_C \, p^\prime$.
These processes necessitate: {\it i)}  high luminosity because they are exclusive; {\it ii)}  high enough $Q^2$, in order to produce the various charmed mesons, and {\it iii)} a wide kinematical range in Bjorken $x$. 

Finally, a few questions have emerged concerning on one side  the applicability of dispersion relations to deeply virtual exclusive processes \cite{Goldstein:2009ks}, and on the other, the commonly assumed partonic picture of the ERBL region \cite{Goldstein:2010ce}. We believe these issues are unique to the exclusive nature of the 
investigated processes and that, in order to establish the correct support, crossing symmetry, and analyticity properties that are necessary to establish dispersion relations, hence a partonic interpretation of GPDs, one needs to include final state interactions/coherent effects  in the proton off-forward structure functions.
Newer deeply virtual exclusive cross section and asymmetry measurements in extended kinematical regimes will provide essential tests of the theory. 

A number of questions remain that are being addressed in ongoing work. The connection between the Regge-like behavior of these GPDs and the more general form of variable mass diquark exchanges has opened up the possibility of having the Regge behavior emerge from diquark mass variation \cite{Goldstein:2010gu}. 
A second, important concern is the inclusion of anti-quarks and gluons, 
whose contribution will affect the low $x_{Bj}$ dependence, particularly the singlet, crossing even GPDs, whose Regge behavior is dominated by Pomeron exchange. 

At this point we see that through the use of physically motivated models and the new horizons provided by the EIC, a far reaching interpretation of the separate spin-dependent GPDs and thereby, a picture of the transverse structure of the nucleons will emerge. 
The connection of chiral odd GPDs to the transversity structure of the nucleon is of great interest as a manifestation of quark and gluon orbital angular momentum.

\subsection{Transverse spin from pseudoscalar meson production}
%Deeply virtual exclusive leptoproduction of photons and mesons (DVCS and DVMP) can be described in terms of Generalized Parton Distributions (GPDs). 
The basic definition of the quark-nucleon GPDs is through off-forward matrix elements of quark field correlators.
%\begin{eqnarray}
%\label{corr1}
%\Phi_{ab} = \int  \, 
%\frac{ d y^-}{2 \pi} \,  
%e^{i y^- X} \, \langle P^\prime S^\prime \mid \overline{\psi}_b(0)  \psi_a(y^-) \mid P S \rangle
%\end{eqnarray}
%where we write the Dirac indices explicitly. 
Contracting with the Dirac matrices, $\gamma^\mu$ or $ \gamma^\mu \gamma^5$ ($\sigma^{\mu \nu} \gamma^5$), and integrating over the internal quark momenta gives rise to the four Chiral even (odd) GPDs, $H, E$ or $\widetilde{H}, \widetilde{E}$ (chiral odd:  $H_T, E_T, \widetilde{H}_T, \widetilde{E}_T$ ~\cite{Diehl:2003ny}). The crucial connection of the 8 GPDs to spin dependent observables in DVCS and DVMP is through the helicity decomposition~\cite{Diehl:2003ny}, where, for example, one of the chiral even and one chiral odd helicity amplitude is given by substituting explicit Dirac helicity state spinors for nucleons to yield
\begin{equation}
A_{++,++}(X,\xi,t)=\frac{\sqrt{1-\xi^2}}{2}(H^q+{\tilde H}^q-\frac{\xi^2}{1-\xi^2}(E^q+{\tilde E}^q)),
\label{Goldstein:chiraleven}
\end{equation}
%\noindent while one of the chiral odd amplitudes is obtained from  Eq.~\ref{oddgpd},
\begin{equation}
A_{++,--}(X,\xi,t)=\sqrt{1-\xi^2}(H_T^q+\frac{t_0-t}{4M^2}{\tilde H}_T^q-\frac{\xi}{1-\xi^2}(\xi E_T^q+{\tilde E}_T^q)).
\label{Goldstein:chiralodd}
\end{equation}

We have constructed a robust model for the GPDs, extending previous work~\cite{Ahmad:2007vw} that is based on the parameterization of diquark spectators and Regge behavior at small $X$. The GPD model parameters are constrained by their relations to PDFs (at $\zeta=0, t=0$)
%$H^q(X,0,0)=f_1^q(X)$, ${\tilde H}^q(X,0,0)=g_1^q(X)$, $H_T^q(X,0,0)=h_1^q(X)$ 
and to nucleon form factors $F_1(t)$,  $F_2(t)$, $ g_A(t)$, $ g_P(t)$ through the first $x$ moments.
% of $H(X,\zeta,t), E(X,\zeta,t), \tilde{H}(X,\zeta,t), \tilde{E}(X,\zeta,t)$, respectively. These are all normalized to the corresponding charge, anomalous moment, axial charge and pseudoscalar ``charge'', respectively. 
For Chiral odd GPDs there are fewer constraints. $H_T(X,0,0)=h_1(X)$ can be fit to the loose constraints in Ref.~\cite{Anselmino:2007fs} - the first moment of $H_T(X,\xi,t)$ is the ``tensor form factor'', called $g_T(t)$. 
%by H\"{a}gler~[\refcite{Haegler}]. Further, 
It is conjectured that the first moment of $2{\tilde H}_T^q(X,0,0)+E_T^q(X,0,0)$ is a  ``transverse anomalous moment'', $\kappa_T^q$, with the latter defined by Burkardt~\cite{Burkardt:2005hp}.

\begin{figure} 
\begin{center}
\includegraphics[width=.40\textwidth]{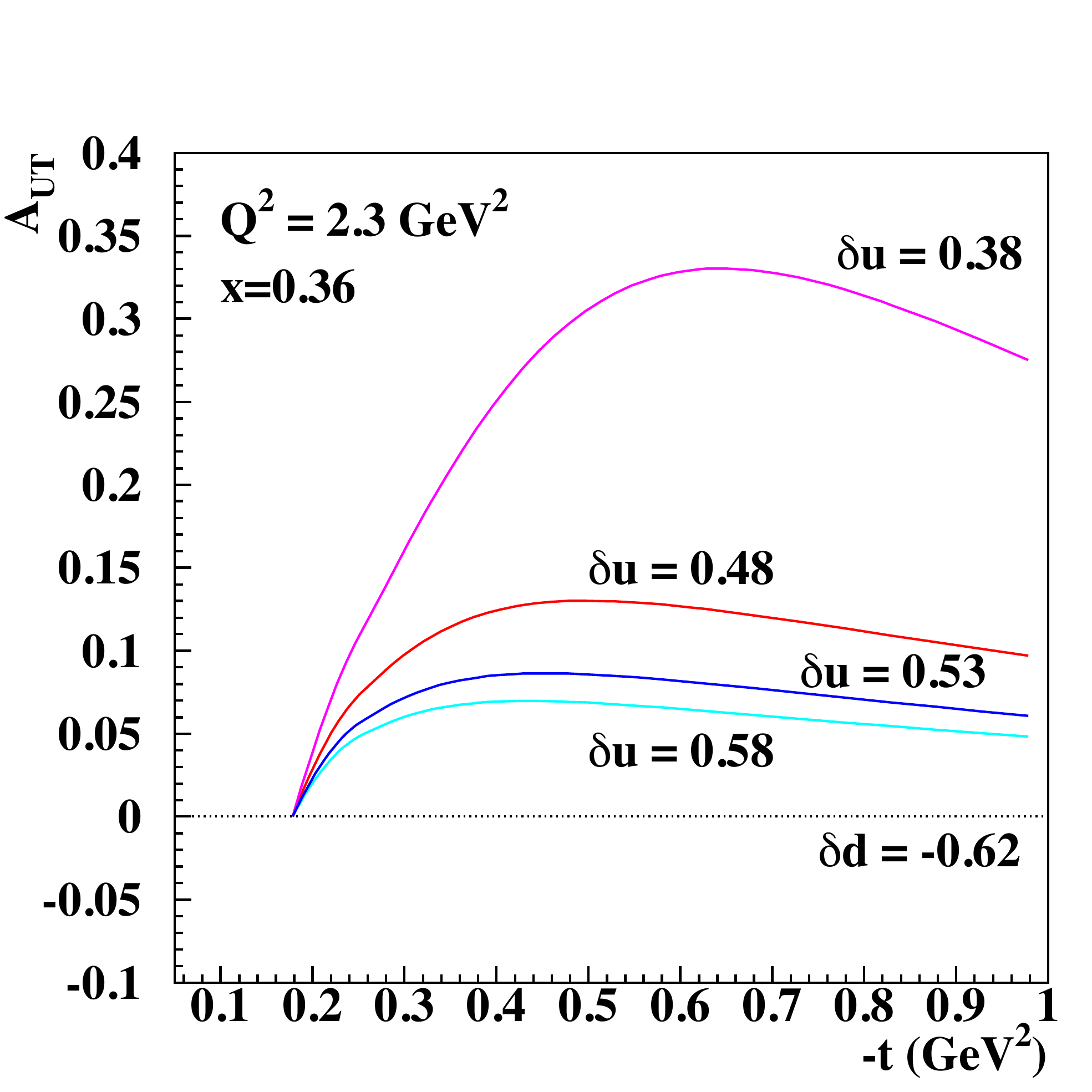} 
\hspace{0.5cm}
\includegraphics[width=.40\textwidth]{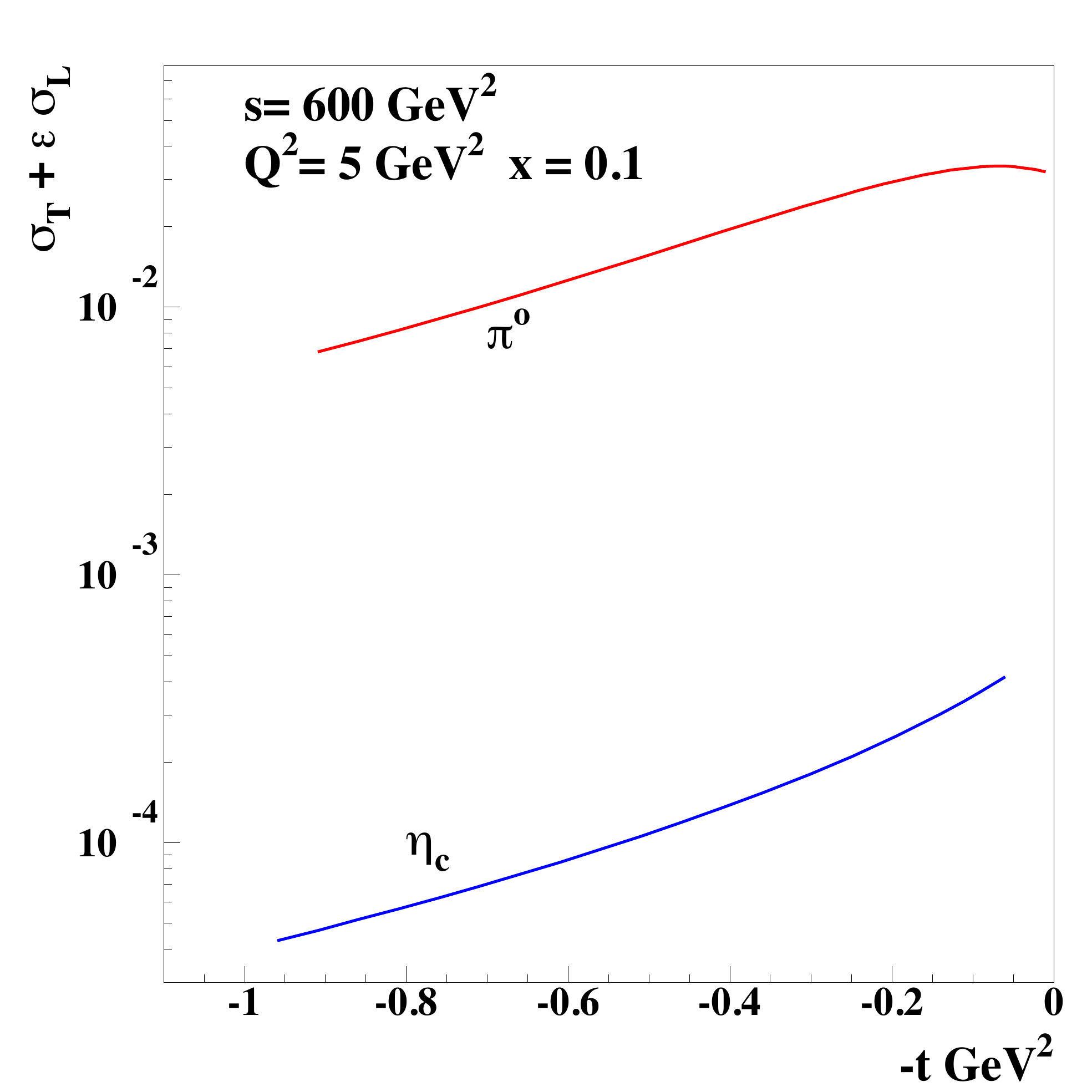} 
\end{center}
%%%%%%%%%%%%%%%%%%%%%%%%%%%%%%
%%%%%%%%%%%%%%%%%%%%%%%%%%%%%%
\caption{\label{fig1} \underline{Left}: Transverse spin asymmetry, $A_{UT}$,  vs. $-t$, at 
$Q^2=2.3$ GeV$^2$, $x_{Bj}=0.36$ for different
values of the tensor charge, $\delta u$, with fixed 
 $\delta d=-0.62$, {\it i.e.} equal to the central 
value extracted in a global fit~\cite{Anselmino:2007fs}.
\underline{Right}: Comparison of $\pi^o$ and $\eta_c$ cross sections. Although $\eta_c$ is the hardest charmed meson 
to detect, the range between the two lines gives an estimate of where the cross sections for the other processes
will lie.}
\end{figure}

With our {\it ansatz} many observables can be determined in parallel with corresponding Regge predictions. Since the initial work~\cite{Ahmad:2008hp}, we have undertaken a more extensive parameterization,and  presented several new predictions~\cite{Goldstein:2010gu}. Here we show one example -  the transversely polarized target asymmetry, in Fig.~\ref{fig1}.
%%%%%%%%%%%%%%%%%%%%%%%%%%%%%%
%%%%%%%%%%%%%%%%%%%%%%%%%%%%%%

%\end{center}

In the case of $\pi^0$ production there are important constraints that restrict the GPDs that contribute. Consider the $t$-channel quantum numbers corresponding to combinations of  GPDs. The $x$ moments of the GPDs have expansions in terms of $t$-dependent form factors and polynomials in $\xi$. It has been shown by Lebed and Ji for pdfs~\cite{Ji:2000id} and Haegler for GPDs~\cite{Hagler:2004yt}
 (see also Ref.~\cite{Diehl:2007jb}), that these moments have $t$-channel angular momentum decompositions, as appropriate for $t$-channel exchanges, as well as Regge poles. 
 From the $t-$channel perspective $\gamma^* + \pi^0$, which has C-parity negative, goes into a $q\bar{q}$ pair 
 %(actually a non-local pair of field operators that have an operator product expansion and Mellin moments), 
 which subsequently becomes an $N\bar{N}$ system. In that chain, each part has the same $J^{PC}$. 

Consider first the chiral even GPDs that contribute to $\pi^0$. 
%In the table we show the relevant crossing (anti-)symmetric combinations of the GPDs. 
The crossing odd $\widetilde{H}$ has contributions from $2^{--}, 4^{--}$ and higher. Although the crossing even , C-even, version has the boundary function $g_1(x)$, we expect that the odd case is supressed. There are several different reasons that this GPD is not expected to contribute at leading order, as we have shown elsewhere \cite{Ahmad:2008hp}. Primarily, there cannot be a $0^{--}$ coupling to $\gamma+\pi^0$ or  $N \bar{N}$. The $2^{--}$ appears in the triplet spin with L=2. 
%For simple resonance exchanges, there would be an angular momentum barrier compared to the $J=1$ exchanges. 
In Regge language the trajectory with $2^{--}$ is non-leading even signature and requires a ``nonsense'' factor to kill the $0^{--}$pole, thereby suppressing the effect in the physical region~\cite{Goldstein:1973xn}. 
%Furthermore, there is no well established $2^{--}$ isoscalar or isovector  ($\rho$ or $\omega$) meson below mass of 2 GeV/c$^2$, although there is a nearly degenerate  pair, $1^{--} \omega(1650)$ with  a $3^{--} \omega(1670)$. Since these can be categorized as $S=1 \, \& \, L=2$, then the  $2^{--}$ would lie in this region of masses. This puts the Regge trajectory going through $J=2$ at $t \approx 2.8 GeV^2$, which will lie well below the $\rho$ and lower than  $a_1$ and $b_1$, minimizing its importance at small $x$ as well.
The crossing odd $\widetilde{E}$ has contributions from $1^{+-}, 2^{--}, 3^{+-},$ etc., so it is the leading candidate for chiral even GPDs that contribute to $\pi^0$.  Its first moment is the pseudoscalar form factor, for 
%\begin{center}
%\begin{table}
%\tbl{Chiral even GPDs \& $J^{PC}$}
%{ \begin{tabular}{| c| c || c |}
%    \hline
%\multicolumn{3}{|c|}{GPDs' $t$-channel $J^{P\, (C \, =\, -)}$  for $\pi^0$ production} \\ \hline
%    crossing symmetrized GPD & $J^{PC}$ & Spin \& (crossing) \\ \hline
%    $H(x,\xi,t)-H(-x,\xi,t)$ & $ 0^{++}, 2^{++}, . . . $ & $S=1$ (odd) \\ \hline
%     $E(x,\xi,t)-E(-x,\xi,t)$ & $ 0^{++}, 2^{++}, . . . $ & $S=1$ (odd) \\ \hline
%    $\widetilde{H}(x,\xi,t)+\widetilde{H}(-x,\xi,t)$ & $1^{++}, 3^{++}, . . .$ & $S=1$ (odd)  \\ \hline
% $\widetilde{E}(x,\xi,t)+\widetilde{E}(-x,\xi,t)$ & $0^{-+},1^{++}, 2^{-+}, . . . $ & $S=0$ (even) \& $1$ (odd) \\ \hline
%   $\widetilde{H}(x,\xi,t)-\widetilde{H}(-x,\xi,t)$ & $2^{--}, 4^{--} . . .$ & $S=1$ (even) \\ \hline       
%   $\widetilde{E}(x,\xi,t)-\widetilde{E}(-x,\xi,t)$ & $1^{+-}, 2^{--}, 3^{+-},  . . .$ & $S=0$ (odd) \& $1$ (even) \\ \hline    $H_T(x,\xi,t)+H_T(-x,\xi,t)$ & $ 1^{+-}, 2^{--}, 3^{+-}, . . . $ & $S=1$ (odd) \\ \hline
%   \end{tabular}}
%    \caption{Some representative GPDs \& $J^{PC=-}$}
 %   \label{table_even}
 %   \end{table}
%\end{center}
which the main contribution is the $\pi$ itself. However, for the neutral $\pi$ this is not the case - there is no $\pi$ pole 
%because there is no $\gamma\rightarrow \pi^0 + \pi^0$ coupling. How does this effect the $\pi^0$ production? It should be said that the role of the charged pion pole in the GPD is not settled in any case, so its contribution to charged $\pi$ production is not agreed upon.
Our approach is that the pole would be included in the GPD, as in the review of Goeke, $et al.$~\cite{Goeke:2001tz}, rather than providing a separate contribution to the amplitudes, as in Ref.~\cite{Goloskokov:2009ia}. That being said, in $\pi^0$ the non-pole contribution to the form factor is relevant, but undoubtedly small~\cite{Goeke:2001tz}. 
%How do we estimate that?

\subsection{$\pi^0$ and Pseudoscalar Production}
The measured cross section for $\pi^0$ is sizable and has large transverse $\gamma^*$ contributions. This indicates that the main contributions should come from chiral odd GPDs, for which the $t$-channel decomposition is richer. In particular, because these GPDs arise from the Dirac matrices $\sigma^{\mu \nu}$, there are 2 series of $J^{PC}$ values for each GPD~\cite{Hagler:2004yt} corresponding to space-space or time-space combinations - $1^{--}$ and $1^{+-}$. These series occur for 3 of the 4 chiral odd GPDs, the exception being $\widetilde{E}_T$. We are thus led to the conclusion that chiral odd GPDs will dominate the neutral pseudoscalar leptoproduction cross sections. 
This result has interesting consequences. For one thing, in a factorized handbag picture, these GPDs will couple to the hard part, the $\gamma^*+\rm{quark} \rightarrow \pi^0 +\rm{quark}$ providing the $\pi^0$ couples through $\gamma^5$, which is naively twist 3, rather than the twist 2 $\gamma^+ \gamma^5$.  Nevertheless, the previous arguments support this choice. Secondly the vector  $1^{--}$ and axial vector $1^{+-}$ in the $t$-channel, viewed as particles ($\rho^0, \omega \,\, \rm{and} \,\,b_1^0, h$), couple primarily to the transverse virtual photon. 
For Reggeons, the $1^{--}$ does not couple at all to the longitudinal photon, while the axial vector $1^{+-}$ does through helicity flip~\cite{Goldstein:1973xn}. Guided by these observations~\cite{Ahmad:2008hp}, we assume the hard part depends on whether the exchange quantum numbers are in the vector or axial vector series, thereby introducing orbital angular momentum into the model. We use $Q^2$ dependent electromagnetic ``transition'' form factors for vector or axial vector quantum numbers going to a pion. We calculate these using PQCD for $q+\bar{q} + \gamma^*(Q^2) \rightarrow q+\bar{q}$ and a standard $z-$dependent pion wave function, convoluted in an impact parameter representation that allows orbital excitations to be easily implemented. 

With our model for the chiral odd, spin-dependent GPDs and these transition form factors, we can obtain the full range of cross sections and asymmetries in kinematic regimes that coincide with ongoing JLab experiments. (A similar emphasis on chiral odd contributions for $\pi$ electroproduction has recently been proposed~\cite{Goloskokov:2009ia}, although the details of that model are quite different from ours.) We are able to predict the important transverse photon contributions to the observables~\cite{Ahmad:2008hp}.  In figure \ref{fig1} (left) we show one striking example of predictions that depend on the values of the tensor charges, thereby providing a means to narrow down those important quantities. This program has been presented~\cite{Goldstein:2010gu} and  further details will soon appear, as the refinements of the chiral odd parameterization are completed \cite{GGL_progress}. In figure \ref{fig1} (right) we show the cross section contribution, $\sigma_T + \epsilon \sigma_L$ for charmed meson production, as compared to $\pi^o$ production \cite{Liuti:2010xy}. 

\subsection{Dispersion Relations}

At the heart our understanding of the role of GPDs in exclusive leptoproduction reactions are the analyticity properties of the amplitudes. We have examined the applicability of Dispersion Relations (DRs) to the GPD formulation of DVCS. Unitarity and completeness are crucial ingredients in establishing analytic properties of the amplitudes. 
The amplitudes are analytic in energy variables, which allows the amplitudes (``Compton Form Factors'' or CFF's) to satisfy DRs relating real and imaginary parts. The imaginary part of a CFF is given by the GPD evaluated at the kinematic point where the returning quark has only transverse momentum relative to the nucleon direction. Then the DR can determine the real part thereby~\cite{Diehl:2007jb} (although this is a region of kinematics that is not easily interpreted in the parton picture). However, at non-zero momentum transfer the DRs require integration over unphysical regions of the variables and that region is considerable, growing larger as $Q^2$ grows. This is reminiscent of the problems with fixed $t$ dispersion relations for elastic scattering that were dealt with in the 1950's, although in the present context the proposed solutions are not relevant. Because of the difficulty of interpolating through the unphysical region,  the real parts must still be measured. Fortunately this can be done for chiral even GPDs by using interference with the Bethe-Heitler contribution~\cite{Goldstein:2009ks}. 

\subsection{Critique of the Naive Partonic Interpretation of GPDs in the ERBL region}

We have also investigated the analyticity in the $X < \zeta$ region (the ERBL region), which conventionally is described as a quark-antiquark or meson distribution in the proton. We are concerned that the completeness relations that lead to analyticity become harder to interpret from the $X=\zeta$ kinematic point to the ERBL kinematics. The formal derivation of the parton model for pdf's in terms of QCD operators is well established~\cite{Jaffe:1983hp}.   We extended the derivation of the parton model from  connected matrix elements for  non-local quark and gluon field operators in inclusive hard processes to the non-forward GPDs~\cite{Goldstein:2010ce}. Without invoking any field theoretic model, such as the Light Front Hamiltonian picture or related models (e.g.~\cite{ Diehl:1998sm,Ji:2006ea}), we are left with the semi-disconnected graphs delineated by Jaffe~\cite{Jaffe:1983hp}.  That is, at leading twist, the kinematics require semi-disconnected amplitudes, {\it i.e.} vacuum fluctuations, that vitiate the partonic interpretation. 

The impasse in trying to give a partonic interpretation of the ERBL region could be overcome by considering multiparton configurations {\it i.e.} extending the definition in terms of quark field correlators to more than two parton fields. 
One suggestive approach is to restore the sensible partonic picture by including gluon exchange, appearing as an initial or final state interaction that ``dresses'' the struck or returning quark. These configurations allow us to describe the ERBL region in terms of connected diagrams, while maintaining a sensible physical picture for the kinematics - the struck quark returns to form the final hadron through the gluonic background, i.e. final state interaction. At first order, the quark is accompanied by a gluon. This work is ongoing, and suggests an important connection between partonic distributions and final state interactions.

\subsection{Acknowledgments}
This work is supported in part by U.S. Department of Energy Grant Nos. DE- 
FG02-92ER40702 (G. R. G.) and DE-FG02-01ER4120 (S. L).

%%%%%%%%%%%%%%%%%%%%%%%%%%%
%%%%%%%%%%%%%%%%%%%%%%%%%%%

\printindex

\end{document}